\documentclass[a4paper]{article}

\usepackage{INTERSPEECH2019}
\newcommand{\etal}{\textit{et al.~}}

\usepackage{tikz}
\usepackage{pgfplots}
\usepackage{xcolor}
\pgfplotsset{grid style={dotted,black}}
\usepackage{subcaption}
\usepackage{cite}

\title{Speech Model Pre-training for End-to-End Spoken Language Understanding}
\name{Loren Lugosch$^1$, Mirco Ravanelli$^1$, Patrick Ignoto$^2$,\\Vikrant Singh Tomar$^2$, Yoshua Bengio$^{1,3}$}
\address{
  $^1$Universit\'e de Montr\'eal / Mila, $^2$Fluent.ai\\$^3$CIFAR Fellow}
\email{\{lugoschl, mirco.ravanelli, yoshua.bengio\}@mila.quebec\\ \{patrick.ignoto, vikrant\}@fluent.ai}

\begin{document}

\maketitle
\begin{abstract}
Whereas conventional spoken language understanding (SLU) systems map speech to text, and then text to intent, end-to-end SLU systems map speech directly to intent through a single trainable model. Achieving high accuracy with these end-to-end models without a large amount of training data is difficult. We propose a method to reduce the data requirements of end-to-end SLU in which the model is first pre-trained to predict words and phonemes, thus learning good features for SLU. We introduce a new SLU dataset, Fluent Speech Commands, and show that our method improves performance both when the full dataset is used for training and when only a small subset is used. We also describe preliminary experiments to gauge the model's ability to generalize to new phrases not heard during training.

\end{abstract}

\section{Introduction}
Spoken language understanding (SLU) systems infer the \textit{meaning} or \textit{intent} of a spoken utterance \cite{renato_SLU}.
This is crucial for voice user interfaces, in which the speaker's utterance needs to be converted into an action or query. 
For example, for a voice-controlled coffee machine, an utterance like ``make me a large coffee with two milks and a sugar, please'' might have an intent representation like \texttt{\{drink: "coffee", size: "large", additions: 
[\{type: "milk", count: 2\}, \{type: "sugar", count: 1\}]\}}.

\begin{figure}
    \centering
    \includegraphics[scale=0.375,trim={0 3cm 0 2cm},clip]{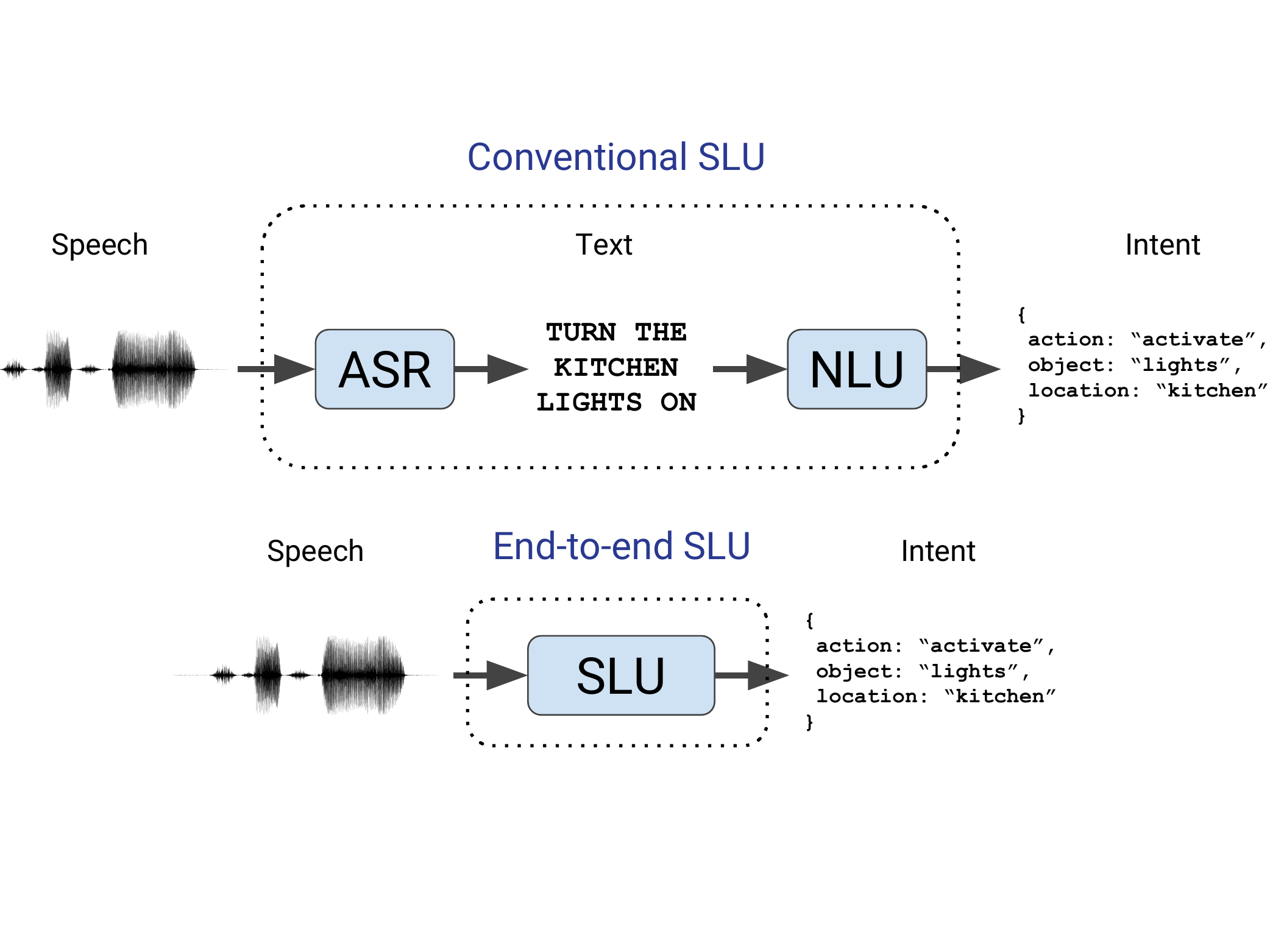}
    \caption{Conventional ASR $\rightarrow$ NLU system for SLU versus end-to-end SLU.}
    \label{fig:SLU_diagram}
\end{figure}

The conventional SLU pipeline is composed of two modules: an automatic speech recognition (ASR) module that maps the speech to a text transcript, and a natural language understanding (NLU) module that maps the text transcript to the speaker's intent \cite{Mesnil2015, Coucke,Gorin1997}. An alternative approach that is beginning to gain popularity is end-to-end SLU \cite{Qian2017, Serdyuk2018, Chen2018, Haghani2018, Renkens2018}. 
In end-to-end SLU, a single trainable model maps the speech audio directly to the speaker's intent without explicitly producing a text transcript (Fig.\ \ref{fig:SLU_diagram}). Unlike the conventional SLU pipeline, end-to-end SLU:
\begin{itemize}
    \item directly optimizes the metric of interest (intent recognition accuracy),
    \item does not waste modeling effort on estimating the text, thus yielding a more compact model and avoiding an error-prone intermediate step involving search algorithms, language models, finite state transducers, etc.,
    \item and enables harnessing aspects of the utterance that may be relevant for inferring the intent, but are not present in the text transcript, such as prosody.
\end{itemize}

End-to-end models have been made possible by deep learning, which automatically learns hierarchical representations of the input signal \cite{Goodfellow-et-al-2016-Book,lecun1998gradient, Krizhevsky2012,Graves2014, Amodei2015}. Speech is natural to represent in a hierarchical way: waveform $\rightarrow$ phonemes $\rightarrow$ morphemes $\rightarrow$ words $\rightarrow$ concepts $\rightarrow$ meaning. However, because speech signals are high-dimensional and highly variable even for a single speaker, training deep models and learning these hierarchical representations without a large amount of training data is difficult.

The computer vision \cite{Yosinski2014, simonyan2014very}, natural language processing \cite{dai2015semi, howard2018universal, Radford, sanh2018hmtl, devlin2018bert}, and ASR \cite{wang2015, IEEEexample:intro1, Kunze, Ghahremani2017} communities have attacked the problem of limited supervised training data with great success by pre-training deep models on related tasks for which there is more training data.
Following their lead, we propose an efficient ASR-based pre-training methodology in this paper and show that it may be used to improve the performance of end-to-end SLU models, especially when the amount of training data is very small.

Our contributions are as follows:
\begin{itemize}
    \item We introduce a dataset for realistic SLU experiments. 
    \item We use this dataset to demonstrate effective speech model pre-training techniques for low-resource SLU.
    \item We make our code\footnote{\texttt{github.com/lorenlugosch/pretrain\_speech\_model}} and data\footnote{\texttt{fluent.ai/research/fluent-speech-commands/}} publicly available for researchers to replicate and build on our work.
\end{itemize}

\section{Related work}

Three key papers describing end-to-end SLU were written by Qian \etal \cite{Qian2017}, Serdyuk \etal \cite{Serdyuk2018}, and Chen \etal \cite{Chen2018}. Serdyuk \etal in \cite{Serdyuk2018} use no pre-training whatsoever. Qian \etal in \cite{Qian2017} use an auto-encoder to initialize the SLU model. Chen \etal \cite{Chen2018} pre-train the first stage of an SLU model to recognize graphemes; the softmax outputs of the first stage are then fed to a classifier second stage. The model proposed in this paper is similar to theirs, but removes the restriction of the softmax bottleneck and uses alternative training targets, as we will describe later.

More recently, Haghani \etal in \cite{Haghani2018} compare four types of sequence-to-sequence models for SLU, including a direct model (end-to-end with no pre-training) and a multi-task model (uses a shared encoder whose output is ingested by a separate ASR decoder and SLU decoder). The model proposed here is somewhat similar to their multi-task model, although we do not use or require the ASR targets during SLU training.

The work listed above deals with very high resource SLU---in \cite{Haghani2018}, for instance, the Google Home \cite{li2017acoustic} dataset consists of 24 million labeled utterances.
In contrast, Renkens \etal in \cite{Renkens2018} consider the problem of end-to-end SLU with limited training data, and find that capsule networks \cite{sabour2017dynamic}, compared to conventional neural network models, are more easily capable of learning end-to-end SLU from scratch.
However, they do not consider the effect of pre-training on other speech data. 

This previous work has all been conducted on datasets that are closed-source or too small to test hypotheses about the amount of data required to generalize well. The lack of a good open-source dataset for end-to-end SLU experiments makes it difficult for most people to perform high-quality, reproducible research on this topic. We therefore created a new SLU dataset, the ``Fluent Speech Commands'' dataset, which Fluent.ai has released along with this paper. 

\section{Dataset}
This section describes the structure and creation of Fluent Speech Commands.

\subsection{Audio and labels}
The dataset is composed of 16 kHz single-channel \texttt{.wav} audio files. Each audio file contains a recording of a single spoken English command that one might use for a smart home or virtual assistant, like ``put on the music'' or ``turn up the heat in the kitchen''.

Each audio is labeled with three \textit{slots}: action, object, and location. A slot takes on one of multiple \textit{values}: for instance, the ``location'' slot can take on the values ``none'', ``kitchen'', ``bedroom'', or ``washroom''. We refer to the combination of slot values as the \textit{intent} of the utterance. The dataset has 31 unique intents in total. We do not distinguish between domain, intent, and slot prediction, as is sometimes done in SLU \cite{wang2005spoken}.

The dataset can be used as a multi-label classification task, where the goal is to predict the action, object, and location labels. Since the slots are not actually independent of each other, a more careful approach would model the relationship between slots, e.g. using an autoregressive model, as in  \cite{Haghani2018}. We use the simpler multi-label classification approach in this paper, so as to avoid the issues sometimes encountered training autoregressive models and instead focus on questions related to generalization using a simpler model. Alternately, the 31 distinct intents can be ``flattened'' and used as 31 distinct labels for a single-label classification task. 

For each intent, there are multiple possible wordings: for example, the intent \texttt{\{action: "activate", object: "lights", location: "none"\}} can be expressed as ``turn on the lights'', ``switch the lights on'', ``lights on'', etc.. These phrases were decided upon before data collection by asking employees at Fluent.ai, including both native and non-native English speakers, for various ways in which they might express a particular intent. 
There are 248 different phrases in total.

\subsection{Data collection}
The data was collected using crowdsourcing. Each speaker was recorded saying each wording for each intent twice. The phrases to record were presented in a random order. Participants consented to data being released and provided demographic information about themselves. The demographic information about these anonymized speakers (age range, gender, speaking ability, etc.) is included along with the dataset.

The data was validated by a separate set of crowdsourcers.
All audios deemed by the crowdsourcers to be unintelligible or contain the wrong phrase were removed. The total number of speakers, utterances, and hours of audio remaining is shown in Table \ref{tab:dataset}. 

\begin{table}[]
  \caption{Information about the Fluent Speech Commands dataset.}
  \label{tab:dataset}
  \centering
  \begin{tabular}{l r r r}
    \toprule
    \textbf{Split} & \textbf{\# of speakers} & \textbf{\# of utterances} & \textbf{\# hours}\\
    \midrule
    Train & 77 & 23,132   & 14.7  \\
    Valid  & 10 & 3,118    & 1.9    \\
    Test  & 10 & 3,793   &  2.4 \\
    \midrule
    Total  & 97 & 30,043  & 19.0 \\
    \bottomrule
  \end{tabular}
  
\end{table}

\subsection{Dataset splits}
The utterances are randomly divided into train, valid, and test splits in such a way that no speaker appears in more than one split. Each split contains all possible wordings for each intent, though our code has the option to include data for only certain wordings for different sets, to test the model's ability to recognize wordings not heard during training. The dataset has a \texttt{.csv} file for each split that lists the speaker ID, file path, transcription, and slots for all the \texttt{.wav} files in that split.

\subsection{Related datasets}
Here we review some related public datasets and show the gap that Fluent Speech Commands fills.

The Google Speech Commands dataset \cite{Warden2018} (to which the name ``Fluent Speech Commands'' is an homage) is a free dataset of 30 single-word spoken commands (``yes'', ``no'', ``stop'', ``go'', etc.). This dataset is suitable for keyword spotting experiments, but not for SLU.

ATIS is an SLU dataset consisting of utterances related to travel planning\footnote{\texttt{https://catalog.ldc.upenn.edu/LDC94S19}}. This dataset can only be obtained expensively from the Linguistic Data Consortium.

The Grabo, Domotica, and Patcor datasets are three related datasets of spoken commands for robot control and card games developed by KU Leuven\footnote{\texttt{https://www.esat.kuleuven.be/psi/spraak/downloads/}} and used in \cite{Renkens2018}. These datasets are free, but have only a small number of speakers and phrases.

The free audio-based Snips SLU Dataset \cite{Saade} is quite close to what we want. This dataset has a great variety of phrases, but less audio---2,946 English utterances and 1,138 French utterances, compared to 30,043 in Fluent Speech Commands.

% The Dialog State Tracking Challenge \cite{henderson2014second} also has utterances with rich SLU annotations. -- supposedly audio also? tried contacting the authors, got "bad address" for all three

In contrast to these datasets, Fluent Speech Commands is simultaneously audio-based, reasonably large, and free, and contains several multiple-word commands corresponding to each of the intents, with multiple recordings for each wording.

\section{Model and Pre-training Strategy}
\begin{figure}
    \centering
    \includegraphics[scale=0.375,trim={0 2cm 0 0cm},clip]{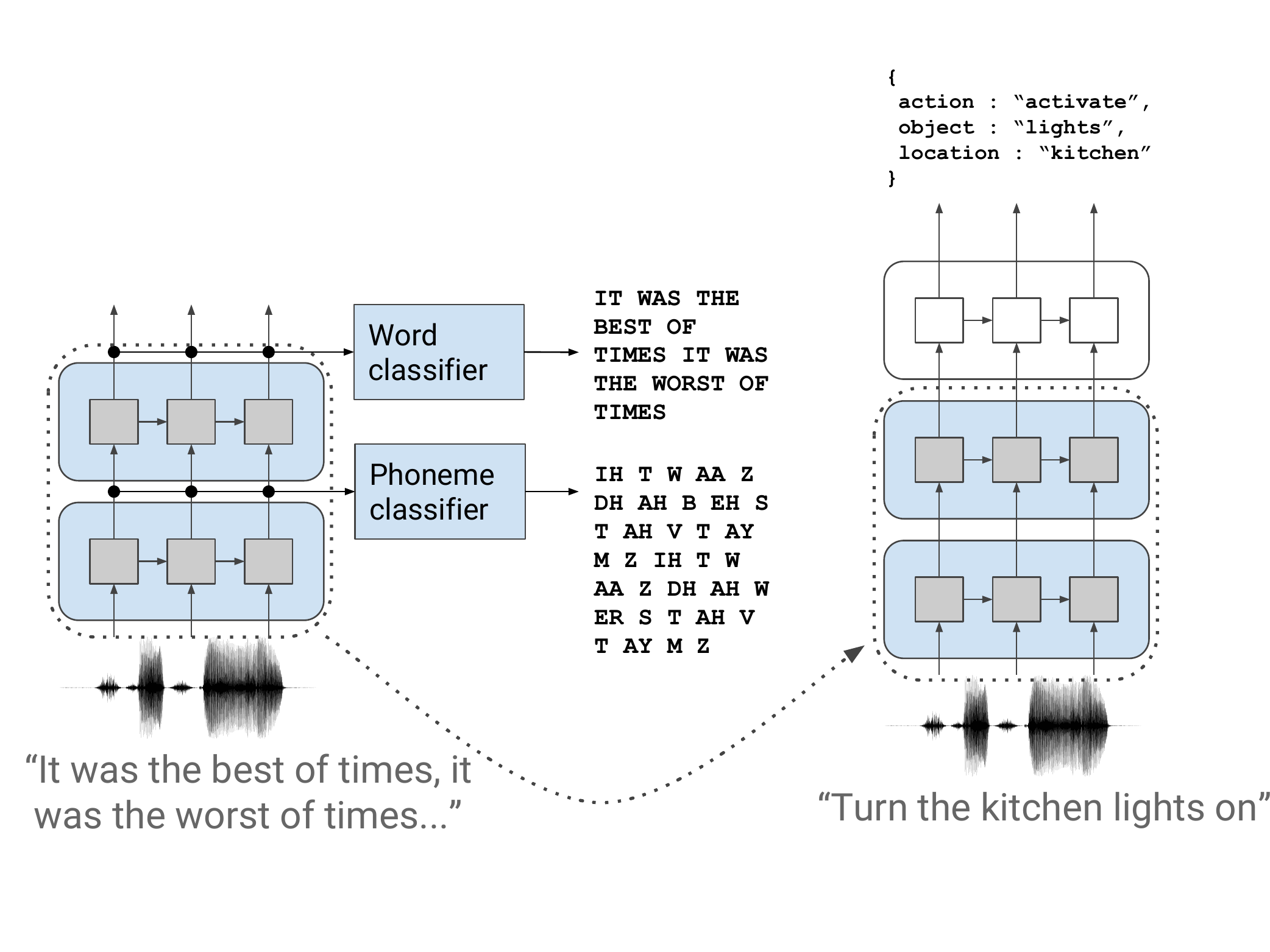}
    \caption{The lower layers of the model are pre-trained using ASR targets (words and phonemes). The word and phoneme classifiers are discarded, and the features from the pre-trained part of the model (blue) are used as the input to the subsequent module (white), which is trained using SLU targets.}
    \label{fig:SLU}
\end{figure}
Using the dataset described in the previous section, we test the performance of our proposed model and pre-training strategy (Fig.\ \ref{fig:SLU}), which we describe here.
The model is a deep neural network consisting of a stack of modules, where the first modules are pre-trained to predict phonemes and words. The word and phoneme classifiers are discarded, and the entire model is then trained end-to-end on the supervised SLU task.
Here we justify and give more details about these design decisions.

\subsection{Which ASR targets to use?}
ASR models are trained using a variety of targets, including phonemes, graphemes, wordpieces, or more recently whole words \cite{45675, Audhkhasi2017, Kartik2018}. We choose whole words as the pre-training targets, since this is what a typical NLU module would expect as input. A large vocabulary ASR dataset contains too many unique words (LibriSpeech \cite{LibriSpeech} has more than 200,000) to efficiently assign an output to each one; we only assign a label to the 10,000 most common words. This leaves much of the pre-training data without any labels, which wastes data. By using phonemes as intermediate pre-training targets \cite{Fernandez2007, sanh2018hmtl, caruana1997multitask}, we are able to pre-train on speech segments with no word label. Additionally, we find that using phonemes as intermediate targets speeds up word-level pre-training \cite{nguyen2019, hierarchical_CTC, krishna2018hierarchical}.

We use the Montreal Forced Aligner \cite{McAuliffe2017} to obtain word- and phoneme-level alignments for LibriSpeech, and we pre-train the model on the entire 960 hours of training data using these alignments\footnote{Our alignments can be downloaded from\\ \texttt{https://zenodo.org/record/2619474\#.XKDP2VNKg1g}}$^,$\footnote{We use textgrid.py to process the alignments.\\ \texttt{https://github.com/kylebgorman/textgrid}}. Using force-aligned labels has the additional benefit of enabling pre-training using short, random crops rather than entire utterances, which reduces the computation and memory required to pre-train the model.

\subsection{Phoneme module}
The first module takes as input the audio signal and outputs $\mathbf{h}^{\text{phoneme}}$, a sequence of hidden representations that are pre-trained to predict phonemes. The phoneme-level logits are computed using a linear classifier:
\begin{equation}
    \mathbf{l}^{\text{phoneme}} = W^{\text{phoneme}}\mathbf{h}^{\text{phoneme}} + b^{\text{phoneme}}.
\end{equation}

The phoneme module is implemented using a SincNet layer \cite{Ravanelli2018,sincnet_irasl}, which processes the raw input waveform, followed by multiple convolutional layers and recurrent layers with pooling (to reduce the sequence length) and dropout. More detailed hyperparameters can be found in our code.

% (explain that pooling reduces sequence length 

% Mirco: You can at least say that you are using bidirectional LSTM or GRU?
\subsection{Word module}
The second module takes as input $\mathbf{h}^{\text{phoneme}}$ and outputs $\mathbf{h}^{\text{word}}$. Similar to the phoneme-level module, it uses recurrent layers with dropout and pooling, and is pre-trained to predict whole word targets using another linear classifier:
\begin{equation}
    \mathbf{l}^{\text{word}} = W^{\text{word}}\mathbf{h}^{\text{word}} + b^{\text{word}}.
\end{equation}

Notice that the input to this module is $\mathbf{h}^{\text{phoneme}}$, not $\mathbf{l}^{\text{phoneme}}$, and likewise the output to the next stage is $\mathbf{h}^{\text{word}}$, not $\mathbf{l}^{\text{word}}$. There are two good reasons for forwarding $\mathbf{h}$ instead of $\mathbf{l}$. The first is that we don't want to remove a degree of freedom from the model: the size of $\mathbf{l}$ is fixed by the number of targets, and this would in turn fix the size of the next layer of the model. The second reason is that computing  $\mathbf{l}^{\text{word}}$ requires multiplying and storing a large ($\approx$ 2.5 million parameters) weight matrix, and by discarding this matrix after pre-training, we save on memory and computation.
% Mirco: One reason is that the output layers is very task-dependent, while if we take a hidden representation before we can have more chances to generalize. This is way is done in speaker verification in the context of d-vectors or x-vectors.
\subsection{Intent module}
\begin{figure*}[t] %[t!]
    \centering
    \begin{subfigure}[b]{0.5\textwidth}
\centering
\begin{tikzpicture}
        \begin{axis}[
            % legend style={at={(0.5,0.3)},anchor=north},
            % height=8cm, %11cm,
            % width=8cm,
            % ymin=0,
            % ymax=1,
            % grid=both,
            % xlabel=Training time (epochs),
            % ylabel=Accuracy
            legend style={at={(0.65,0.45)},anchor=north},
            height=6cm,
            width=8cm,
            ymin=0,
            ymax=1,
            grid=both,
            xlabel=Training time (epochs),
            ylabel=Accuracy
        ]
        
        \addplot[color=black] coordinates {
            (1, 0.467607)
            (2, 0.683772)
            (3, 0.757216)
            (4, 0.774214)
            (5, 0.813663)
            (6, 0.800513)
            (7, 0.841886)
            (8, 0.839320)
            (9, 0.859205)
            (10, 0.813983)
            (11, 0.844452)
            (12, 0.850866)
            (13, 0.808852)
            (14, 0.847659)
            (15, 0.846055)
            (16, 0.862733)
            (17, 0.864978)
            (18, 0.870430)
            (19, 0.837716)
            (20, 0.859205)
        };
        \addlegendentry{No pre-training}
        
        \addplot[color={rgb:red,0;green,0;blue,1}] coordinates {
            (1, 0.816870)
            (2, 0.871071)
            (3, 0.880372)
            (4, 0.882938)
            (5, 0.898974)
            (6, 0.897691)
            (7, 0.901860)
            (8, 0.905388)
            (9, 0.905709)
            (10, 0.908595)
            (11, 0.906030)
            (12, 0.914047)
            (13, 0.910520)
            (14, 0.904426)
            (15, 0.910840)
            (16, 0.911161)
            (17, 0.910199)
            (18, 0.910840)
            (19, 0.912444)
            (20, 0.912765)
        };
        \addlegendentry{No unfreezing}
        
        \addplot[color=violet] coordinates {
            (1, 0.816870)
            (2, 0.892880)
            (3, 0.907312)
            (4, 0.913085)
            (5, 0.929442)
            (6, 0.928159)
            (7, 0.932970)
            (8, 0.925914)
            (9, 0.926235)
            (10, 0.931046)
            (11, 0.934253)
            (12, 0.936818)
            (13, 0.935536)
            (14, 0.932970)
            (15, 0.939384)
            (16, 0.940026)
            (17, 0.936498)
            (18, 0.932008)
            (19, 0.926876)
            (20, 0.936177)
        };
        \addlegendentry{Unfreeze word layers}
        
        \addplot[color=red] coordinates {
            (1, 0.816870)
            (2, 0.892880)
            (3, 0.907312)
            (4, 0.913727)
            (5, 0.918538)
            (6, 0.922386)
            (7, 0.908916)
            (8, 0.886786)
            (9, 0.885183)
            (10, 0.897049)
            (11, 0.903143)
            (12, 0.888069)
            (13, 0.902502)
            (14, 0.905067)
            (15, 0.894804)
            (16, 0.901860)
            (17, 0.891597)
            (18, 0.892239)
            (19, 0.890956)
            (20, 0.889994)
        };
        \addlegendentry{Unfreeze all layers}
        
        \end{axis}
        
    \end{tikzpicture}

\caption{Full dataset.}
\label{fig:full_fig}
    \end{subfigure}%
    ~ 
    \begin{subfigure}[b]{0.5\textwidth}
        \centering
\begin{tikzpicture}
        % \begin{semilogyaxis}[
        \begin{axis}[
            legend style={at={(0.65,0.45)},anchor=north},
            height=6cm,
            width=8cm,
            ymin=0,
            ymax=1,
            grid=both,
            xlabel=Training time (epochs),
            ylabel=Accuracy
        ]
        
        \addplot[color=black] coordinates {
            (0, 0.034958)
(1, 0.061578)
(2, 0.076331)
(3, 0.163887)
(4, 0.197242)
(5, 0.260103)
(6, 0.305324)
(7, 0.370750)
(8, 0.406992)
(9, 0.455420)
(10, 0.496151)
(11, 0.504169)
(12, 0.569275)
(13, 0.549070)
(14, 0.592367)
(15, 0.616100)
(16, 0.608082)
(17, 0.655548)
(18, 0.627004)
(19, 0.663566)
(20, 0.641758)
(21, 0.653945)
(22, 0.694035)
(23, 0.671905)
(24, 0.676395)
(25, 0.720654)
(26, 0.712316)
(27, 0.709429)
(28, 0.701090)
(29, 0.695959)
(30, 0.707826)
(31, 0.705260)
(32, 0.711674)
(33, 0.702053)
(34, 0.702694)
(35, 0.694035)
(36, 0.696280)
(37, 0.723861)
(38, 0.707826)
(39, 0.687300)
(40, 0.729634)
(41, 0.710391)
(42, 0.706222)
(43, 0.719051)
(44, 0.733162)
(45, 0.744387)
(46, 0.745350)
(47, 0.754009)
(48, 0.737332)
(49, 0.708146)
(50, 0.720975)
(51, 0.761065)
(52, 0.737332)
(53, 0.710712)
(54, 0.705260)
(55, 0.724824)
(56, 0.725144)
(57, 0.739577)
(58, 0.745029)
(59, 0.744708)
(60, 0.740539)
(61, 0.733804)
(62, 0.718089)
(63, 0.714240)
(64, 0.705581)
(65, 0.704298)
(66, 0.675433)
(67, 0.712636)
(68, 0.749840)
(69, 0.744067)
(70, 0.746312)
(71, 0.743425)
(72, 0.762989)
(73, 0.759461)
(74, 0.740218)
(75, 0.748877)
(76, 0.744708)
(77, 0.747274)
(78, 0.727069)
(79, 0.747274)
(80, 0.702053)
(81, 0.717447)
(82, 0.765555)
(83, 0.753688)
(84, 0.745670)
(85, 0.751764)
(86, 0.747915)
(87, 0.728993)
(88, 0.720334)
(89, 0.721616)
(90, 0.735728)
(91, 0.742463)
(92, 0.738294)
(93, 0.758499)
(94, 0.751123)
(95, 0.755933)
(96, 0.762348)
(97, 0.768441)
(98, 0.766838)
(99, 0.751123)
        };
        \addlegendentry{No pre-training}
        
        \addplot[color={rgb:red,0;green,0;blue,1}] coordinates {
(0, 0.232200)
(1, 0.567992)
(2, 0.660359)
(3, 0.688262)
(4, 0.721937)
(5, 0.734766)
(6, 0.739577)
(7, 0.753368)
(8, 0.761706)
(9, 0.789609)
(10, 0.810776)
(11, 0.820398)
(12, 0.822001)
(13, 0.837075)
(14, 0.830340)
(15, 0.836434)
(16, 0.841886)
(17, 0.838999)
(18, 0.846697)
(19, 0.847017)
(20, 0.849904)
(21, 0.855356)
(22, 0.860808)
(23, 0.861129)
(24, 0.864336)
(25, 0.864657)
(26, 0.854073)
(27, 0.861450)
(28, 0.857601)
(29, 0.858242)
(30, 0.863374)
(31, 0.863695)
(32, 0.860808)
(33, 0.861770)
(34, 0.863374)
(35, 0.860487)
(36, 0.859205)
(37, 0.862733)
(38, 0.856639)
(39, 0.865619)
(40, 0.871713)
(41, 0.866581)
(42, 0.860167)
(43, 0.865940)
(44, 0.866581)
(45, 0.861770)
(46, 0.859205)
(47, 0.861129)
(48, 0.864336)
(49, 0.866581)
(50, 0.864978)
(51, 0.866902)
(52, 0.864015)
(53, 0.864015)
(54, 0.868826)
(55, 0.867543)
(56, 0.865619)
(57, 0.870430)
(58, 0.868185)
(59, 0.866581)
(60, 0.873637)
(61, 0.869468)
(62, 0.860167)
(63, 0.871071)
(64, 0.863374)
(65, 0.865298)
(66, 0.873316)
(67, 0.871713)
(68, 0.871071)
(69, 0.871392)
(70, 0.867543)
(71, 0.864657)
(72, 0.866260)
(73, 0.869468)
(74, 0.867223)
(75, 0.867223)
(76, 0.870109)
(77, 0.865940)
(78, 0.857601)
(79, 0.868185)
(80, 0.868826)
(81, 0.868505)
(82, 0.861129)
(83, 0.861450)
(84, 0.868505)
(85, 0.860167)
(86, 0.865619)
(87, 0.873637)
(88, 0.868185)
(89, 0.868185)
(90, 0.873316)
(91, 0.869468)
(92, 0.865298)
(93, 0.875241)
(94, 0.862733)
(95, 0.862412)
(96, 0.869147)
(97, 0.872675)
(98, 0.869147)
(99, 0.865940)
        };
        \addlegendentry{No unfreezing}
        
        \addplot[color=violet] coordinates {
            (0, 0.232200)
(1, 0.617704)
(2, 0.714881)
(3, 0.756575)
(4, 0.794740)
(5, 0.836434)
(6, 0.847659)
(7, 0.847979)
(8, 0.863053)
(9, 0.875561)
(10, 0.860808)
(11, 0.863695)
(12, 0.867223)
(13, 0.877165)
(14, 0.872354)
(15, 0.873637)
(16, 0.880693)
(17, 0.874599)
(18, 0.881976)
(19, 0.878768)
(20, 0.869147)
(21, 0.877486)
(22, 0.882938)
(23, 0.889994)
(24, 0.883579)
(25, 0.887107)
(26, 0.891918)
(27, 0.884221)
(28, 0.884862)
(29, 0.885183)
(30, 0.885504)
(31, 0.882296)
(32, 0.869468)
(33, 0.876523)
(34, 0.876844)
(35, 0.880693)
(36, 0.878448)
(37, 0.888390)
(38, 0.885824)
(39, 0.881334)
(40, 0.882617)
(41, 0.875882)
(42, 0.884221)
(43, 0.885504)
(44, 0.890314)
(45, 0.882296)
(46, 0.875241)
(47, 0.879731)
(48, 0.881013)
(49, 0.876523)
(50, 0.879410)
(51, 0.881655)
(52, 0.875561)
(53, 0.883258)
(54, 0.882296)
(55, 0.881976)
(56, 0.889031)
(57, 0.884862)
(58, 0.891597)
(59, 0.884862)
(60, 0.893201)
(61, 0.887428)
(62, 0.895767)
(63, 0.892880)
(64, 0.885824)
(65, 0.882296)
(66, 0.887107)
(67, 0.887428)
(68, 0.887107)
(69, 0.889994)
(70, 0.888390)
(71, 0.886145)
(72, 0.884541)
(73, 0.884541)
(74, 0.895767)
(75, 0.891276)
(76, 0.886145)
(77, 0.885504)
(78, 0.884862)
(79, 0.887107)
(80, 0.884541)
(81, 0.876844)
(82, 0.893201)
(83, 0.897691)
(84, 0.890314)
(85, 0.894484)
(86, 0.897370)
(87, 0.895767)
(88, 0.892239)
(89, 0.886466)
(90, 0.884862)
(91, 0.890635)
(92, 0.893842)
(93, 0.890314)
(94, 0.881976)
(95, 0.884862)
(96, 0.891918)
(97, 0.894163)
(98, 0.888069)
(99, 0.887749)
        };
        \addlegendentry{Unfreeze word layers}
        
        \addplot[color=red] coordinates {
(0, 0.232200)
(1, 0.617704)
(2, 0.714881)
(3, 0.754009)
(4, 0.805645)
(5, 0.822322)
(6, 0.833868)
(7, 0.805965)
(8, 0.820398)
(9, 0.820398)
(10, 0.860167)
(11, 0.865940)
(12, 0.851828)
(13, 0.840603)
(14, 0.831944)
(15, 0.850545)
(16, 0.845414)
(17, 0.831302)
(18, 0.854073)
(19, 0.839962)
(20, 0.851187)
(21, 0.863695)
(22, 0.855677)
(23, 0.828736)
(24, 0.837396)
(25, 0.840924)
(26, 0.815587)
(27, 0.811738)
(28, 0.856318)
(29, 0.854394)
(30, 0.867543)
(31, 0.858563)
(32, 0.850545)
(33, 0.858884)
(34, 0.848621)
(35, 0.859205)
(36, 0.872354)
(37, 0.854394)
(38, 0.871071)
(39, 0.870109)
(40, 0.863374)
(41, 0.870750)
(42, 0.876203)
(43, 0.886466)
(44, 0.883258)
(45, 0.850545)
(46, 0.854073)
(47, 0.862412)
(48, 0.849583)
(49, 0.836754)
(50, 0.804362)
(51, 0.797627)
(52, 0.827133)
(53, 0.790892)
(54, 0.837075)
(55, 0.850545)
(56, 0.855035)
(57, 0.858563)
(58, 0.857922)
(59, 0.806286)
(60, 0.838037)
(61, 0.825529)
(62, 0.870430)
(63, 0.863053)
(64, 0.853432)
(65, 0.855677)
(66, 0.872033)
(67, 0.870750)
(68, 0.855677)
(69, 0.863053)
(70, 0.863695)
(71, 0.842848)
(72, 0.844452)
(73, 0.856639)
(74, 0.844452)
(75, 0.847017)
(76, 0.853111)
(77, 0.854394)
(78, 0.842527)
(79, 0.858242)
(80, 0.836113)
(81, 0.851828)
(82, 0.843489)
(83, 0.823926)
(84, 0.846376)
(85, 0.843489)
(86, 0.843810)
(87, 0.813342)
(88, 0.798268)
(89, 0.820718)
(90, 0.795702)
(91, 0.831302)
(92, 0.835151)
(93, 0.835792)
(94, 0.840282)
(95, 0.858563)
(96, 0.855035)
(97, 0.857922)
(98, 0.852470)
(99, 0.862412)
        };
        \addlegendentry{Unfreeze all layers}
        
        \end{axis}
        
    \end{tikzpicture}

\caption{10\% of the dataset.}
\label{fig:subset_fig}
    \end{subfigure}
    \caption{Accuracy on the validation set over time for models trained on (a) the full SLU dataset or (b) 10\% of the dataset.}
\end{figure*}
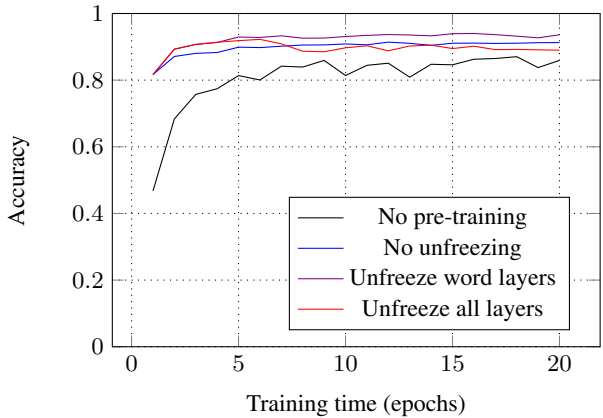
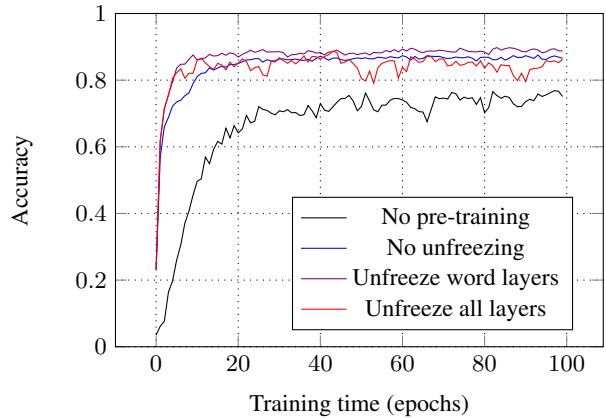
The third module, which is not pre-trained, maps $\mathbf{h}^{\text{word}}$ to the predicted intent.  Depending on the structure of the intent representation, the intent module might take on various forms. Since in this work we use a fixed three-slot intent representation, we implement this module using a recurrent layer, followed by max-pooling to squash the sequence of outputs from the recurrent layer into a single vector of logits corresponding to the different slot values, similar to \cite{Serdyuk2018}. 

\subsection{Unfreezing schedule}
Although the pre-trained model works well as a frozen feature extractor, it may be preferable to ``unfreeze'' its weights and finetune them for the SLU task with backpropagation. Similar to ULMFiT \cite{howard2018universal}, we find that gradually unfreezing the pre-trained layers works better than unfreezing them all at once. 
% Mirco: Interesting, I an try it for my work on self-superbvised learning as well.
We unfreeze one layer each epoch, and stop at a pre-determined layer, which is a hyperparameter.

\section{Experiments}
Here we report results for three experiments on Fluent Speech Commands: using the full dataset, using a subset of the dataset, and using a subset of wordings.

\subsection{Full dataset}
We first trained models given the entire SLU training set. The models used one of: 1) no pre-training (randomly initialized), 2) pre-training with no unfreezing, 3) gradually unfreezing only the word layers, or 4) gradually unfreezing both the word layers and phoneme layers. What we report here as ``accuracy'' refers to the accuracy of all slots for an utterance taken together---that is, if the predicted intent differs from the true intent in even one slot, the prediction is deemed incorrect.

The validation accuracy of these models over time is shown in Fig. \ref{fig:full_fig}. The best results are obtained when only the word layers of the pre-trained model are unfrozen. This may be because the fully unfrozen model begins to forget the more general phonetic knowledge acquired during pre-training. For the test set, the frozen model and partially unfrozen model perform roughly equally well (Table \ref{tab:test_results}, ``full'' column), possibly because the test set is ``easier'' than the validation set. In all cases, the pre-trained models outperform the randomly initialized model. 

\begin{table}[]
  \caption{Accuracy on the test set for different models, given the full training dataset or a 10\% subset of the training data.}
  \label{tab:test_results}
  \centering
  \begin{tabular}{l r r}
    \toprule
    \textbf{Model} & \textbf{Accuracy (full)} & \textbf{Accuracy (10\%)} \\
    \midrule
    No pre-training & 96.6\% & 88.9\% \\
    No unfreezing  & 98.8\% & 97.9\% \\
    Unfreeze word layers  & 98.7\%      & 97.9\% \\
    Unfreeze all layers  & 97.2\%  & 95.8\%   \\
    \bottomrule
  \end{tabular}
  
\end{table}
\subsection{Partial dataset}
To simulate a smaller dataset, we randomly selected 10\% of the training set, and used this instead of the entire training set. Fig. \ref{fig:subset_fig} shows the validation accuracy (on the entire validation set, not a subset) over time. A similar trend is observed as for the entire dataset: unfreezing the word layers works best. The gap in final test accuracy between the randomly initialized model and the pre-trained models increases (Table \ref{tab:test_results}, ``10\%'' column); the final test accuracy for the pre-trained models drops only slightly, further highlighting the advantage of our proposed method.

\subsection{Generalizing to new wordings}
What happens if new wordings appear in the test data that never appear in the training data? This is an important question, since it is generally impractical to try to imagine every possible wording for a particular intent while gathering training data. 

To test this, we trained models on three specific phrases, ``turn on the lights'', ``turn off the lights'', and ``switch on the lights'' (273 utterances total), and tested on those same phrases, as well as a new phrase: ``switch off the lights''. If the model incorrectly infers that utterances that contain ``switch'' always correspond to turning on the lights, it will incorrectly guess that ``switch off the lights'' corresponds to turning on the lights; if the model infers that the presence of the word ``off'' corresponds to turning off the lights, it will generalize to the new phrase. The randomly initialized model was unable to fit this tiny training set, even with a very low learning rate and no regularization. The pre-trained models were able to generalize to the new wording (with 97\% accuracy on the validation set, which contains more examples of the new phrase than of the training phrases).

However, there are many situations in which our model does not correctly generalize. For example, if the model is trained only with examples containing ``bedroom'' and ``washroom'', but then tested on an example containing ``bathroom'', it will guess the intent corresponding to ``bedroom'' because ``bedroom'' sounds more similar to ``bathroom'' than to ``washroom'', even though ``washroom'' is the correct meaning. In text-based NLU, this scenario can be handled using word embeddings, which represent words in such a way that words with similar meanings have similar vector representations \cite{Mesnil2015, mikolov2013efficient}. It may be possible to teach the pre-trained part of the model to output ``embedding-like'' word representations so that the intent module can recognize the meaning of phrases with synonyms.

\section{Conclusion}
In this paper, we proposed a pre-training methodology for end-to-end SLU models, introduced the Fluent Speech Commands dataset, and used this dataset to show that our pre-training techniques improve performance both for large and small SLU training sets. In the future, we plan to continue using Fluent Speech Commands to explore the limitations of end-to-end SLU, like new wordings and synonyms not observed in the SLU dataset, to see if these limitations can be overcome.

\section{Acknowledgements}

We would like to acknowledge the following for
research funding and computing support: NSERC, Calcul Qu\'{e}bec, Compute Canada,
the Canada Research Chairs, and CIFAR. 

Thanks to Dima Serdyuk and Kyle Kastner at Mila, and Farzaneh Fard, Luis Rodriguez Ruiz, Sam Myer, 	
Mohamed Mhiri, and Arash Rad at Fluent.ai for helpful discussions with us about this work.

\bibliographystyle{IEEEtran}

\bibliography{mybib}

\end{document}